\documentclass[fleqn,10pt]{article}
\usepackage[affil-it]{authblk}
\usepackage[english]{babel}
\usepackage{blindtext}
\date{}

\providecommand{\keywords}[1]{\textbf{\textit{Keywords:}} #1}
\providecommand{\msc}[1]{\textbf{\textit{2010 MSC:}} #1}

\usepackage{graphicx}
\usepackage{amsmath}
\usepackage{amsthm}
\usepackage{amssymb}
\usepackage{amsfonts}%for \mathfrak{g}
\usepackage[hidelinks]{hyperref}
\usepackage{booktabs}%for using \toprule, \bottomrule, \midrule, place it before longtable
\usepackage{longtable}
\usepackage{cite}
\usepackage{mathrsfs}

\usepackage[title,titletoc,toc]{appendix}
\newtheorem{thm}{Theorem}[section]
\newtheorem{lem}[thm]{Lemma}
\newtheorem{rem}{Remark}

\newtheorem{case}{Case}
\newtheorem{Case}{Case}
\usepackage{chngcntr}
\counterwithin{case}{section}
\counterwithin{Case}{subsection}
\counterwithin{exmp}{section}
\counterwithin{rem}{section}
\counterwithin{red}{section}
\newtheorem{defi}{Definition}
\usepackage[margin=1in,left=1in,right=1in]{geometry}
\usepackage{subcaption}
\usepackage{float}
\usepackage{placeins}%for float barrier
\allowdisplaybreaks
\title{On invariant analysis and conservation laws for degenerate coupled multi-KdV equations for multiplicity $l = 3$}

\author[1]{R.K. Gupta}
\author[2]{Manjit Singh\thanks{Corresponding Author, Email: manjitcsir@gmail.com.}}

\affil[1]{%
    Centre for Mathematics and Statistics School of Basic and Applied Sciences, Central University of Punjab Bathinda-151001, Punjab, India.}
\affil[2]{%
    Yadavindra College of Engineering Punjabi University Guru Kashi Campus Talwandi Sabo-151302, Punjab, India.}

\begin{document}

\maketitle
\begin{abstract}
The degenerate coupled multi-KdV equations for coupled multiplicity $l=3$ are studied. The equations also known as three fields Kaup-Boussinesq equations are considered for invariant analysis and conservation laws. The classical Lie's symmetry method is used to analyze the symmetries of equations. Based on Killing's form which is invariant of adjoint action, the full classification for Lie algebra is presented. Further, one-dimensional optimal group classification is used to obtain invariant solutions. Besides this,  using general theorem proved by Ibragimov we find several nonlocal conservation laws for these equations. The conserved currents obtained in this work can be useful for the better understanding  of some physical phenomena modeled by the underlying equations.
\end{abstract}
\\%this command can be re-defined from line 6
\msc{70G65,70M60,35L65}
%\pacs{02.20.Sv, 02.30.Jr, 04.20.Jb, 11.30.−j.}%this command can be re-defined from line 7

\section{Introduction}
Recently, degenerate coupled Korteweg-de Vries equations for coupled multiplicity $l= 2, 3, 4$ have been considered for traveling wave solutions by G$\ddot{\rm u}$rses and Pekcan \cite{gurses2014traveling,asli}. The multi-system of Kaup-Boussinesq equations is
given by (see refs. \cite{antonowicz1987coupled,antonowicz1987family})
\begin{equation}{\label{KBQ44}}
\begin{aligned}
u_{t} =&\, \frac{3}{2}\,uu_{x}+q_{x}^{2},\\
q_{t}^{2} =& \,q^{2}u_{x}+\frac{1}{2}\,uq_{x}^{2}+q_{x}^{3},\\
\vdots\\
q_{t}^{l-1} =& \,q^{l-1}u_{x}+\frac{1}{2}\,uq_{x}^{l-1}+v_{x},\\
v_{t}=&\,-\frac{1}{4}u_{xxx}+vu_{x}+\frac{1}{2}\,uv_{x},
\end{aligned}
\end{equation}
where $q^{1} = u$ and $q^{l} = v$. For $l=2$, the system \eqref{KBQ44} reduces to
\begin{equation}{\label{KBQ45}}
\begin{aligned}
u_{t} = &\,\frac{3}{2}\,uu_{x}+v_{x},\\
v_{t} = &\,-\frac{1}{4}\,u_{xxx}+vu_{x}+\frac{1}{2}\,uu_{x}.
\end{aligned}
\end{equation}
The system \eqref{KBQ45} has been studied in detail by G$\ddot{\rm u}$rses and Pekcan \cite{gurses2014traveling} for traveling wave solutions, and they have proved that there exists no asymptotically vanishing traveling
wave solution of system \eqref{KBQ45}. Wazwaz \cite{wazwaz2015generalized} has studied generalized version of system \eqref{KBQ45} for multiple-soliton solutions. It is pertinent to mention here that the system \eqref{KBQ45} which is also known as Kaup-Boussinesq system exhibit same shallow water wave characteristics in the same approximation as the well-known Boussinesq equation in the
lowest order in small parameters controlling weak dispersion and nonlinearity effects \cite{kaup1975higher,el2001integrable,ivanov2009two,ivanov2012integrable}. Moreover, the function $v(x, t)$ denotes the height of the water surface above a horizontal
bottom, whereas the function $u(x, t)$ denotes its velocity averaged over depth. The Kaup-Boussinesq system \eqref{KBQ45} corresponds to the case when the gravity force dominates over the capillary
one and it is completely integrable \cite{kamchatnov2001asymptotic,kamchatnov2003asymptotic,bhrawy2013integrable}.

For $\l = 3$, the system \eqref{KBQ44} has the following form:
\begin{equation}{\label{KBQ1}}
\begin{aligned}
&u_{t}-\frac{3}{2}uu_{x}-v_{x} = 0,\\
&v_{t}-vu_{x}-\frac{1}{2}uv_{x}-w_{x} = 0,\\
&w_{t}+\frac{1}{4}u_{xxx}-wu_{x}-\frac{1}{2}uw_{x} = 0.
\end{aligned}
\end{equation}
The system of equations \eqref{KBQ1} which are also known as three field Kaup-Boussinesq equations \cite{gurses2013integrable} have been discussed for traveling wave solutions \cite{asli},  wherein authors have given general approach to solve equations \eqref{KBQ44} for $l\geq 3$. Subsequently,  by using bifurcation analysis, Li and Chen \cite{li2016bifurcations} have given complete parametric representations of traveling wave solutions of system \eqref{KBQ44} for $l = 2, 3, 4$, which was earlier missing in work of G$\ddot{\rm u}$rses and Pekcan \cite{gurses2014traveling,asli}. In literature, we have noticed that the equations \eqref{KBQ1} have not been completely  analyzed, so in this work, we propose to investigate equations \eqref{KBQ1} for Lie's symmetry analysis and for conservation laws using recently proposed new theorem by Ibragimov.\\
\indent The paper is planned as follow. In \autoref{KBQsec2}, based on classical Lie symmetry analysis we have obtained four-dimensional Lie algebra. Starting with a brief discussion about classification techniques,  Lie algebra is then classified into mutually conjugate classes by identifying Killing's form which is invariant of full adjoint action. Reductions are also presented corresponding to every conjugate class and exact solutions are also obtained. In \autoref{KBQsec3}, based on new theorem proposed by Ibragimov, several nonlocal conservation laws are also constructed. Finally, in \autoref{KBQsec4}, the conclusion is drawn.
\section{Lie symmetry analysis of Kaup-Boussinesq equations \eqref{KBQ1}}{\label{KBQsec2}}
In order to identify Lie point symmetries for Eq. \eqref{KBQ1}, we follow standard procedure given in \cite{ovsi,anco}. So we consider one parameter local Lie group of point transformations,
\begin{equation}{\label{KBQ2}}
\begin{aligned}
&\tilde{x} = x+\epsilon\,\xi(x,t,u,v,w)+O(\epsilon^{2}),\\
&\tilde{t} = t+\epsilon\,\tau(x,t,u,v,w)+O(\epsilon^{2}),\\
&\tilde{u} = u+\epsilon\,\eta_{1}(x,t,u,v,w)+O(\epsilon^{2}),\\
&\tilde{v} = v+\epsilon\,\eta_{2}(x,t,u,v,w)+O(\epsilon^{2}),\\
&\tilde{w} = w+\epsilon\,\eta_{3}(x,t,u,v,w)+O(\epsilon^{2}),
\end{aligned}
\end{equation}
where $\epsilon$ being group parameter. The invariance of Eq. \eqref{KBQ1} under symmetry transformations \eqref{KBQ2} give rise to overdetermined system of linear partial differential equations in $\xi, \tau, \eta_{1}, \eta_{2}$ and  $\eta_{3}$. Such overdetermined system may be derived by considering the associated vector field, which may be expressed as
\begin{align}
\label{KBQ3}V=\xi\frac{\partial}{\partial x}+\tau\frac{\partial}{\partial t}+\eta_{1}\frac{\partial}{\partial u}+\eta_{2}\frac{\partial}{\partial v}+\eta_{3}\frac{\partial}{\partial w}.
\end{align}
Third order prolongation of vector field \eqref{KBQ3} when applied in following manner
\begin{align}
\label{KBQ4}V^{(3)}(\Delta)\big|_{\eqref{KBQ1}}=0,\; \text{here $\Delta$ is system \eqref{KBQ1}},
\end{align}
will give infinitesimals of symmetry transformation as
\begin{eqnarray}{\label{KBQ6}}
\begin{aligned}
&\xi = -\frac{5c_{3}}{6}t+\frac{3c_{1}}{5}x+c_{4},\,\tau = c_{1}t+c_{2},\\
&\eta_{1} = -\frac{2c_{1}}{5}u+c_{3},\;\eta_{2} = -\frac{4c_{1}}{5}v-\frac{2c_{3}}{3}u,\;\eta_{3} =
-\frac{c_{3}}{3}v-\frac{6c_{1}}{5}w.
\end{aligned}
\end{eqnarray}
Infinitesimals \eqref{KBQ6} gives four dimensional Lie algebra
\begin{equation}{\label{KBQ5}}
\begin{aligned}
&V_{1} = \frac{\partial}{\partial t},\;V_{2} = \frac{\partial}{\partial x},\;\text{(translation)}\\
&V_{3} = -\frac{5t}{6}\frac{\partial}{\partial x}+\frac{\partial}{\partial u}-\frac{2u}{3}\frac{\partial}{\partial v}-\frac{v}{3}\frac{\partial}{\partial w},\;\text{(Galilean boost)}\\
&V_{4} = \frac{3x}{5}\frac{\partial}{\partial x}+t\frac{\partial}{\partial t}-\frac{2u}{5}\frac{\partial}{\partial u}-\frac{4v}{5}\frac{\partial}{\partial v}-\frac{6w}{5}\frac{\partial}{\partial w}\,\text{(dilation)}.
\end{aligned}
\end{equation}
The non-zero Lie commutations of Lie algebra \eqref{KBQ5}
\begin{align}
\label{KBQ7}[V_{1}, V_{3}] = -\frac{5V_{2}}{6},\;[V_{1}, V_{4}] = V_{1}, \;[V_{2}, V_{4}] = \frac{3V_{2}}{5},\;[V_{3}, V_{4}] = -\frac{2V_{3}}{5}.
\end{align}
The non-zero Lie brackets \eqref{KBQ7} shows that the Lie algebra \eqref{KBQ5} is solvable.
\subsection{Construction of optimal system for Lie algebra \eqref{KBQ5}}
In symmetry analysis, it is well-known fact that, whenever PDEs or system of PDEs admits symmetry group (or group of invariant transformations), then one can find group invariant solution corresponding to each sub-group by reducing the number of independent variables in the original system. There exist infinitely many such sub-groups and hence infinitely many group invariant solutions. But most of those group invariant solutions would be equivalent by some transformation in full symmetry group. In order to minimize the search of inequivalent group invariant solutions under transformations in full symmetry group, the concept of the optimal system is introduced. Although the classification of Lie algebras by use of adjoint transformations was known to Lie himself, but Ovsiannikov \cite{ovsi} was first to use Lie group classification to derive inequivalent group invariant solutions. Ovsiannikov used a global adjoint matrix to construct optimal systems and he further extended his technique to derive multi-dimensional optimal systems. In the construction of the two-dimensional optimal system, Galas \cite{galas1991exact} made some modifications in the technique of Ovsiannikov by selecting elements from normalizer of the one-dimensional optimal system.

  Apart from these techniques, the method of classifying the sub-algebras proposed by Patera \emph{et al.}  \cite{patera1975continuous} is par excellence (see ref. \cite{gupta2017group} for recent applications) and in their subsequent work \cite{patera1977subalgebras} they have classified all real Lie algebras of dim $\leq$ 4 under the group of inner automorphisms. It is worth mentioning that Lie algebra of  dimension greater than 4 have also been fully classified. For example, Turkowski \cite{turkowski1990solvable} has classified all six dimensional solvable Lie algebras containing four-dimensional nilradical. He has also classified all Lie algebras of dimension up to 9.  Chou \emph{et al.} \cite{chou2001note,chou2004optimal} suggested a slightly modified technique of classifying Lie algebras. They have constructed different varieties of invariants of the group of inner automorphism including numerical and conditional invariants. Despite the early work \cite{olverbook} on group classification by using adjoint actions(or identification of equivalence classes based on sign of Killing's form), the work of Chou \emph{et al.} is very useful as their additional invariants helps to  confirm optimality \emph{i.e.} their  technique  confirms the completeness and mutual inequivalence of representatives of sub-algebras. In present work, we shall use only Killing's form as invariant for detecting all the representatives of sub-algebras. Before going further in this subsection, we shall introduce some definitions and lemmas.
\begin{defi}{\label{KBQdef1}}
\normalfont
The symmetric bilinear form $\varphi$ on the space of Lie algebra $\mathscr{L}$, that is, the mapping $\varphi:\;\mathscr{L}\times\mathscr{L}\rightarrow \mathbb{R} $ is called \emph{invariant} (relative to group $Int\;\mathscr{L}$) if for any inner automorphism $A\in Int\;\mathscr{L}$, and for any $V_{1}, V_{2}\in \mathscr{L}$
\begin{align*}
\varphi\langle A\langle V_{1}\rangle, A\langle V_{2}\rangle\rangle = \varphi\langle V_{1}, V_{2} \rangle.
\end{align*}
In term of adjoint representation $Ad_{g}$, the real function $\varphi$ on Lie algebra $\mathscr{L}$ is invariant if and only if $\varphi(Ad_{g}(V)) = \varphi(V)$ for all $V\in\mathscr{L}$ and $g\in G$ (group generated by $\mathscr{L}$).
\end{defi}
\begin{defi}{\label{KBQdef2}}
\normalfont
Let $\mathscr{L} $ be a Lie algebra, and $V\in\mathscr{L}$. Then the adjoint transformation defined by $V$ is the linear transformation $ad(V):\mathscr{L}\rightarrow\mathscr{L}$ defined by
\begin{align*}
ad(V)(W) = [V, W],\;\text{for all}\; W\in\mathscr{L},
\end{align*}
here $[\cdot\,,\cdot]$ is usual Lie bracket. The exponential of $ad(X)$, usually denoted by $Ad(X)$, is a Lie algebra isomorphism. The symmetric bilinear form $K:\mathscr{L}\times \mathscr{L}\rightarrow \mathbb{R}$ defined by
\begin{align*}
K\langle V, W\rangle = tr(ad(V)\circ ad(W))
\end{align*}
is called {Killing's} form. This Killing's form is invariant of group of inner automorphism $Int \mathscr{L}$, the importance of which we shall realize during construction of optimal system of subalgebras.
\end{defi}
\begin{lem}
\normalfont
Let $V = \sum_{i=1}^{4}a_{i}V_{i}$ be the general element of Lie algebra $\mathscr{L}^{4}$ given by \eqref{KBQ5} and $a_{1},\dots,a_{4}\in \mathbb{R}$. The invariant function $\varphi$ is of the form $f(a_{4})$, here $f$ is arbitrary function.
\begin{proof}
\normalfont
The general invariant function $\varphi$ can be obtained by solving system of linear partial differential equations given by
\begin{align}
\label{KBQ8}a_{4}\frac{\partial \varphi}{\partial a_{2}} = 0,\;a_{4}\frac{\partial \varphi}{\partial a_{1}}-\frac{5a_{3}}{6}\frac{\partial \varphi}{\partial a_{2}} = 0,\;-\frac{2a_{4}}{5}\frac{\partial \varphi}{\partial a_{3}}+\frac{5a_{1}}{6}\frac{\partial \varphi}{\partial a_{2}}= 0,\;-a_{1}\frac{\partial \varphi}{\partial a_{1}}+\frac{2a_{3}}{5}\frac{\partial \varphi}{\partial a_{3}}-\frac{3a_{2}}{5}\frac{\partial \varphi}{\partial a_{2}} = 0.
\end{align}
A straightforward solution of system \eqref{KBQ8} is $f(a_{4})$, for arbitrary function $f$. The procedure for construction of system of PDEs \eqref{KBQ8} is discussed in detail in Ref. \cite{hu2015direct}
\end{proof}
\end{lem}
\begin{lem}{\label{KBQlem}}
\normalfont
The Killing's form is particular invariant which can be derived from general solution of system \eqref{KBQ8}, for Lie algebra \eqref{KBQ5} $K\langle V, V\rangle = \frac{38}{25}a_{4}^{2}$.
\begin{proof}
The direct computations shows that
\begin{align}
ad (V) = \left[ \begin {array}{cccc} -a_{{4}}&0&0&a_{{1}}\\ \noalign{\medskip}
\frac{5a_{3}}{6}&-\frac{3a_{4}}{5}&-\frac{5a_{1}}{6}&\frac{3a_{2}}{5}
\\ \noalign{\medskip}0&0&\frac{2a_{4}}{5}&-\frac{2a_{3}}{5}
\\ \noalign{\medskip}0&0&0&0\end {array} \right].
\end{align}
By definition of Killing's form,  $K\langle V, V\rangle = tr(ad(V)\circ ad(V)) = \frac{38}{25}a_{4}^{2}$.
\end{proof}
\end{lem}
In search of group invariant solutions, one ought to be careful of instances where two group invariant solutions can be recovered from each other by some transformation in full symmetry group. For example, the two group invariant solutions $\Psi_{1}$ and $\Psi_{2}$ are called as essentially inequivalent if it is impossible to connect these solutions by some four parameter group transformation $\tilde{\psi} = \exp\left[\sum_{i=1}^{4}a_{i}V_{i}\right]\psi$. In this manner, the group invariant solutions separate into equivalence classes, and the collection of generators corresponding these classes would constitute an optimal system. In order to find such equivalence classes we define adjoint operator
\begin{align}
\label{KBQ9}\text{Ad}_{\exp(\epsilon V)}\left(W\right) = \exp(-\epsilon \,V)W\exp(\epsilon \,V) = \tilde{W}\left(\epsilon\right)
\end{align}
The adjoint transformation \eqref{KBQ9} can be written through Lie brackets using Campbell-Hausdorff formula as
\begin{align}
\label{KBQ10}\text{Ad}_{\exp(\epsilon V)}\left(W\right) = W-\epsilon [V,W]+\frac{\epsilon^{2}}{2}[V,[V,W]]-\dots,
\end{align}
where $[. , .]$ is Lie bracket defined by \eqref{KBQ7}. Let $V = \sum_{i=1}^{4}a_{i}V_{i}$, based on Lie brackets defined at \eqref{KBQ7} and formula \eqref{KBQ10}, straightforward calculations  shows that
\begin{align}{\label{KBQ11}}
 &\text{Ad}_{\exp(\epsilon_{3} v_{3})}\,\text{Ad}_{\exp(\epsilon_{4} v_{4})}\,\text{Ad}_{\exp(\epsilon_{1} v_{1})}\, \text{Ad}_{\exp(\epsilon_{2} v_{2})}(V)=\,\sum_{i=1}^{4}\tilde{a}_{i}V_{i}.
\end{align}
The full adjoint transformation \eqref{KBQ11} in matrix notation
\begin{align}{\label{KBQ12}}
A = \left[ \begin {array}{cccc} {{\rm e}^{\epsilon_{{4}}}}&0&0&-\epsilon_
{{1}}{{\rm e}^{\epsilon_{{4}}}}\\ \noalign{\medskip}-\frac{5\epsilon_{{3}}}{6}\,{{\rm e}^{
\epsilon_{{4}}}}&{{\rm e}^{\frac{3\epsilon_{{4}}}{5}}}&\frac{5\epsilon_{{1}}}{6}\,
{{\rm e}^{\frac{3\epsilon_{{4}}}{5}}}&-\frac{3\epsilon_{{2}}}{5}\,{
{\rm e}^{\frac{3\epsilon_{{4}}}{5}}}+\frac{5\epsilon_{{1}}\epsilon_{{3}}}{6}\,{{\rm e}^{\epsilon_{
{4}}}}\\ \noalign{\medskip}0&0&{{\rm e}^{-\frac{2\epsilon_
{{4}}}{5}}}&\frac{2\epsilon_{{3}}}{5}\\ \noalign{\medskip}0&0&0&1\end {array}
 \right].
\end{align}
The construction of adjoint matrix \eqref{KBQ12} is discussed in ref. \cite{ovsi}, where the coefficients $\tilde{a}_{1},\dots,\tilde{a}_{4}$ in \eqref{KBQ11} are given by
\begin{equation}{\label{KBQ13}}
\begin{aligned}
\tilde{a}_{1}& = -a_{{4}}\epsilon_{{1}}{{\rm e}^{\epsilon_{{4}}}}+a_{{1}}{{\rm e}^{
\epsilon_{{4}}}},\\
\tilde{a}_{2}& = \frac{5a_{{3}}\epsilon_{{1}}}{6}\,{{\rm e}^{\frac{3\epsilon_{{4}}}{5}}}-\frac{3a_{{4}}
\epsilon_{{2}}}{5}\,{{\rm e}^{\frac{3\epsilon_{{4}}}{5}}}+a_{{2}}{{\rm e}^{\frac{3\epsilon_{{4}}}{5}}}+\frac{5a_{{4}}\epsilon_{{1}}\epsilon_{{3}}}{6}\,{{\rm e}^{\epsilon_{{4}}}}
-\frac{5a_{{1}}\epsilon_{{3}}}{6}\,{{\rm e}^{\epsilon_{{4}}}},\\
\tilde{a}_{3}& = \frac{2a_{{4}}\epsilon_{{3}}}{5}+a_{{3}}{{\rm e}^{-\frac{2\epsilon_{{4}}}{5}}},\\
\tilde{a}_{4}& = a_{4}.
\end{aligned}
\end{equation}
The last equation in \eqref{KBQ13} agrees with invariance of Killing's form under full adjoint transformation \eqref{KBQ11}.
\begin{thm}{\label{KBQthm}}
\normalfont
The one dimensional optimal system corresponding to Lie algebra \eqref{KBQ6} is
$\left\{V_{1}, V_{2}, V_{3}, V_{4}, \alpha\,V_{1}\pm V_{3}\right\}$.
\begin{proof}
Let $V = \sum_{i=1}^{4}a_{i}V_{i}$ and $K = \frac{38}{25}a_{4}^{2}$. We have following cases for $K$
\begin{Case}
\normalfont
For $K\neq 0$, we take $a_{4} = 1$. Choosing $\epsilon_{4} = 0$, system of equations \eqref{KBQ13} becomes
\begin{align*}
\tilde{a}_{1}& = -\epsilon_{{1}}+a_{{1}},\\
\tilde{a}_{2}& = \frac{5a_{{3}}\epsilon_{{1}}}{6}-\frac{3
\epsilon_{{2}}}{5}+a_{{2}}+\frac{5\epsilon_{{1}}\epsilon_{{3}}}{6}\,
-\frac{5a_{{1}}\epsilon_{{3}}}{6},\\
\tilde{a}_{3}& = \frac{2\epsilon_{{3}}}{5}+a_{{3}},\\
\tilde{a}_{4}& = 1.
\end{align*}
The selection $\epsilon_{{1}}=a_{{1}},\epsilon_{{2}}={\frac {25\,a_{{3}}a_{{
1}}}{18}}+\frac{5a_{{2}}}{3},\epsilon_{{3}}=-\frac{5a_{{3}}}{2}
$, gives $\tilde{a}_{1} = \tilde{a}_{2} = \tilde{a}_{3} = 0$, we obtain simplification $V = V_{4}$
\end{Case}
\begin{Case}
\normalfont
For $K = 0$, we have to take $a_{4} = 0$.
\begin{enumerate}
\item [(1)]$a_{3} = 1$. Choosing $\epsilon_{{4}}=\frac{5}{2}\,\ln  \left( a_{{3}} \right) $ gives $\tilde{a}_{3} = \pm 1$, and appropriate selection of  $\epsilon_{1}, \epsilon_{3}$ gives $\tilde{a}_{1} = a_{{1}}{a_{{3}}}^{\frac{5}{2}}$ and $\tilde{a}_{2} = 0$. We obtain simplification $V = \alpha\,V_{1}\pm V_{3}$, $\alpha = a_{{1}}{a_{{3}}}^{\frac{5}{2}}$
    \item [(2)] $a_{3} = 0$. The system \eqref{KBQ13} reduce to $\tilde{a}_{1} = a_{{1}}{{\rm e}^{\epsilon_{{4}}}}, \tilde{a}_{2} = a_{{2}}{{\rm e}^{\frac{3\epsilon_{{4}}}{5}}}-\frac{5a_{{1}}\epsilon_{{3}}}{6}\,{{\rm e}^{\epsilon_
{{4}}}}
$. By taking $\epsilon_{{3}}=\,{\frac {6 a_{{2}}{{\rm e}^{-\frac{2\epsilon_{{4}}}{5}}}}{5a
_{{1}}}}
$, we obtain simplification $V = V_{1}$.
\item[(3)] $a_{3} = 0, a_{1} = 0$. In this case we obtain straightforward simplification $V = V_{2}$.
\item[(4)]$a_{3}\neq 0, a_{1} = 0$. $\tilde{a}_{2} = \frac{5a_{{3}}\epsilon_{{1}}}{6}\,{{\rm e}^{\frac{3\epsilon_{{4}}}{5}}}+a_{{2}}{
{\rm e}^{\frac{3\epsilon_{{4}}}{5}}}
, \tilde{a}_{3} = a_{{3}}{{\rm e}^{-\frac{2\epsilon_{{4}}}{5}}}$. By taking $\epsilon_{{1}}=-\,{\frac {6a_{{2}}}{5a_{{3}}}}$ we obtain simplification $V = V_{3}$.
\end{enumerate}
\end{Case}
\end{proof}
\end{thm}
\subsection{Symmetry reductions and invariant solutions}
By virtue of vector fields $V_{1}$ and $V_{2}$, we can see that the equations \eqref{KBQ1} admits symmetry in space and time translation. So by letting $\xi = x-c\,t$,  we have similarity transformations $u = F(\xi), v = G(\xi)$ and $w = H(\xi)$. Substituting into system \eqref{KBQ1} we obtain
\begin{subequations}
\begin{align}
\label{KBQ122}&-c\,F_{{\xi}}-\frac{3}{2}\,FF_{{\xi}}-G_{{\xi}}=0,\\
\label{KBQ133}&-c\,G_{{\xi}}-GF_{{\xi}}-\frac{1}{2}\,FG_{{\xi}}-H_{{\xi}}=0,\\
\label{KBQ14}&-c\,H_{{\xi}}+\frac{1}{4}\,F_{{\xi,\xi,\xi}}-HF_{{\xi}}-\frac{1}{2}\,FH_{{\xi}}=0.
\end{align}
\end{subequations}
Integrating \eqref{KBQ122} wrt $\xi$ gives
\begin{align}
\label{KBQ15}G = -c\,F-\frac{3}{4}\,F^{2}+d_{1},\;\text{here}\, d\,\text{is constant of integration}.
\end{align}
Again substituting this $G$ into \eqref{KBQ133} gives
\begin{align*}
H_{\xi} = (c^{2}-d_{1})F_{\xi}+3c\,FF_{\xi}+\frac{3}{2}F^{2}F_{\xi},
\end{align*}
integrating once wrt $\xi$
\begin{align}
\label{KBQ16}H = (c^{2}-d_{1})F+\frac{3c}{2}F^{2}+\frac{1}{2}F^{3}+d_{2}.
\end{align}
Substituting \eqref{KBQ15} and \eqref{KBQ16} into \eqref{KBQ14} gives
\begin{align}{\label{KBQ17}}
-{c}^{3}\,F_{{\xi}}+cd_{{1}}\,F_{{\xi}}-\frac{9}{2}\,{c}^{2}FF_{{\xi}}-\frac{9}{2}\,cF_{{
\xi}}{F}^{2}+\frac{1}{4}\,F_{{\xi,\xi,\xi}}+\frac{3}{2}d_{{1}}\,F_{{\xi}}F-\frac{5}{4}\,F_{{
\xi}}{F}^{3}-d_{{2}}\,F_{{\xi}}
=0.
\end{align}
Integrating \eqref{KBQ17} wrt $\xi$ and then second integration after using integrating factor $F_{\xi}$ gives
\begin{align}
\label{KBQ18}(F_{\xi})^{2} = \frac{1}{2}\,F^{5}+3c\,F^{4}+(6c^{2}-2d_{1})\,F^{3}+4(c^{3}-cd_{1}+d_{2})\,F^{2}+8d_{3}\,F+8d_{4},
\end{align}
here $c, d_{1}, d_{2}, d_{3}, d_{4}$ are constants of integration.
Detailed discussion about solution of \eqref{KBQ18} can be seen in ref. \cite{asli}.
The reductions corresponding to rest of vectors fields have been classified in following cases.
\begin{Case}
\normalfont
Reduction under subalgebra $V_{3}$.
\begin{itemize}
\item Similarity variables.
\begin{eqnarray}{\label{KBQ46}}
\begin{aligned}\xi = &\,t\\u=&\,\,F \left( t \right) -\,{\frac {6x}{5t}}\\v=&\,G \left( t \right) +\,{\frac {4ux}{5t}}+{\frac {12\,{x}^{2}}{25\,{t}^
{2}}}
\\w=&\,H \left( t \right) +\,{\frac {2vx}{5t}}-{\frac {4\,u{x}^{2}}{25\,{t}^
{2}}}-{\frac {8\,{x}^{3}}{125\,{t}^{3}}}
\end{aligned}
\end{eqnarray}
\item Reduced system. Substituting \eqref{KBQ46} into \eqref{KBQ1}, the reduced system is obtained as follows:
    \begin{eqnarray}{\label{KBQ47}}
    \begin{aligned}
    &tF_{{t}}+F=0\\&2t\,{F}^{2}-5{t}^{2}\,G_{{t}}-4tx\,F_{{t}}-4x\,F-4t\,G=0\\&4tx\,{F}^{2}+5{t}^{2}\,FG-25{t}^{3}\,H_{{t}}-10{t}^{2}x\,G_{{t}}-4t{x}^{2}\,F
_{{t}}\\&-4{x}^{2}\,F-8tx\,G-30{t}^{2}\,H=0
\end{aligned}
\end{eqnarray}
\item Similarity solutions.
\begin{eqnarray}
\begin{aligned}
u=&\,{\frac {c_{{1}}}{t}}-\,{\frac {6x}{5t}}\\v=&\,-\,{\frac {{c_{{1}}}^{2}}{3{t}^{2}}}+{\frac {c_{{2}}}{{t}^{4/5}}}+
\,{\frac {4c_{{1}}x}{5{t}^{2}}}-{\frac {12\,{x}^{2}}{25\,{t}^{2}}}\\w=&\,-\,{\frac {c_{{1}}c_{{2}}}{3{t}^{9/5}}}+{\frac {{c_{{1}}}^{3
}}{27{t}^{3}}}+{\frac {c_{{3}}}{{t}^{6/5}}}-\,{\frac {2x{c_{{1}}}^{2}
}{15{t}^{3}}}\\&+{\frac {2xc_{{2}}}{5{t}^{9/5}}}+{\frac {4\,c_{{1}}{x}^{
2}}{25\,{t}^{3}}}-{\frac {8\,{x}^{3}}{125\,{t}^{3}}}
\end{aligned}
\end{eqnarray}

\end{itemize}
\end{Case}

\begin{Case}{\label{KBQCaseKB2}}
\normalfont
Reduction under subalgebra $V_{4}$.
\begin{itemize}
\item Similarity variables.
\begin{eqnarray}{\label{KBQ48}}
\begin{aligned}
\xi=&\frac{t}{x^{\frac{5}{3}}},\;u=\,\frac{1}{x^{\frac{2}{3}}}F(\xi),\;v=\,\frac{1}{x^{\frac{4}{3}}}G(\xi),\,
     w=\,\frac{1}{x^{2}}H(\xi)
     \end{aligned}
     \end{eqnarray}
     \item Reduced system. Substituting \eqref{KBQ48} into \eqref{KBQ1}, the reduced system is obtained as follows:
         \begin{eqnarray}{\label{KBQ49}}
         \begin{aligned}
         &15\,F\xi\,F_{{\xi}}+6\,{F}^{2}+10\,\xi\,G_{{\xi}}+8\,G+6\,F_{{\xi}}=0\\&5\,F\xi\,G_{{\xi}}+10\,G\xi\,F_{{\xi}}+8\,FG+10\,\xi\,H_{{\xi}}+12\,H+
6\,G_{{\xi}}=0\\&125\,{\xi}^{3}F_{{\xi,\xi,\xi}}-90\,F\xi\,H_{{\xi}}-180\,H\xi\,F_{{\xi
}}+750\,{\xi}^{2}F_{{\xi,\xi}}\\&-180\,FH+830\,\xi\,F_{{\xi}}+80\,F-108\,
H_{{\xi}}=0
\end{aligned}
\end{eqnarray}
\item Similarity solutions.
\begin{eqnarray}{\label{KBQ50}}
\begin{aligned}
u=\,\frac{1}{x^{\frac{2}{3}}}\sum_{n = 0}^{\infty}P_{n}\,\left(\frac{t}{x^{\frac{5}{3}}}\right)^{n},
\;v=\,\frac{1}{x^{\frac{4}{3}}}\sum_{n = 0}^{\infty}Q_{n}\,\left(\frac{t}{x^{\frac{5}{3}}}\right)^{n},\;w=\,\frac{1}{x^{2}}\sum_{n = 0}^{\infty}R_{n}\,\left(\frac{t}{x^{\frac{5}{3}}}\right)^{n},
\end{aligned}
\end{eqnarray}
where the coefficients $P_{n}, Q_{n}$ and $R_{n}$ are obtained in following Theorem \ref{KBQthmKB1}.
\end{itemize}
\end{Case}

\begin{Case}{\label{KBQCaseKB3}}
\normalfont
Reduction under subalgebra $\alpha\,V_{1}+V_{3}$.
\begin{itemize}
\item Similarity variables.
\begin{eqnarray}{\label{KBQ51}}
\begin{aligned}
\xi=&\,{\frac {12\,\alpha\,x}{5}}+{t}^{2}\\u=&\,-F \left( \xi \right) -{\frac {t}{\alpha}}\\v=&\,G \left( \xi \right) -\,{\frac {2tF \left( \xi \right)}{3\alpha}}+\,{\frac {4x}{5\alpha}}\\w=&\,H \left( \xi \right) -\,{\frac {t\,G \left( \xi \right) }{3\alpha}}-{
\frac {4x\,F \left( \xi \right) }{15\,\alpha}}-{\frac {4\,tx}{45\,{
\alpha}^{2}}}-{\frac {2\,t\xi}{27\,{\alpha}^{3}}}
\end{aligned}
\end{eqnarray}
\item Reduced system. Substituting \eqref{KBQ51} into \eqref{KBQ1}, the reduced system is obtained as follows:
    \begin{eqnarray}
    \begin{aligned}
    &18\,F{\alpha}^{2}F_{{\xi}}+12\,{\alpha}^{2}G_{{\xi}}-1=0\\&9\,F{\alpha}^{3}G_{{\xi}}+18\,G{\alpha}^{3}F_{{\xi}}+18\,{\alpha}^{3}H
_{{\xi}}+4\,\alpha\,\xi\,F_{{\xi}}+6\,F\alpha=0\\&3888\,{\alpha}^{6}F_{{\xi,\xi,\xi}}-1350\,F{\alpha}^{4}H_{{\xi}}-2700
\,H{\alpha}^{4}F_{{\xi}}+450\,F{\alpha}^{2}\xi\,F_{{\xi}}\\&+150\,{F}^{2}
{\alpha}^{2}-375\,G{\alpha}^{2}-125\,\xi=0
\end{aligned}
\end{eqnarray}
\item Similarity solutions.
\begin{eqnarray}
\begin{aligned}
u=&\,-\sum_{n = 0}^{\infty}P_{n}\,\left({\frac {12\,\alpha\,x}{5}}+{t}^{2}\right)^{n} -{\frac {t}{\alpha}},\\v=&\,\sum_{n = 0}^{\infty}Q_{n}\,\left({\frac {12\,\alpha\,x}{5}}+{t}^{2}\right)^{n} -\,{\frac {2t}{3\alpha}}\sum_{n = 0}^{\infty}P_{n}\,\left({\frac {12\,\alpha\,x}{5}}+{t}^{2}\right)^{n}+\,{\frac {4x}{5\alpha}},\\w=&\,\sum_{n = 0}^{\infty}R_{n}\,\left({\frac {12\,\alpha\,x}{5}}+{t}^{2}\right)^{n} -\,{\frac {t\, }{3\alpha}}\sum_{n = 0}^{\infty}Q_{n}\,\left({\frac {12\,\alpha\,x}{5}}+{t}^{2}\right)^{n}\\&-{
\frac {4x }{15\,\alpha}}\sum_{n = 0}^{\infty}P_{n}\,\left({\frac {12\,\alpha\,x}{5}}+{t}^{2}\right)^{n}-{\frac {4\,tx}{45\,{
\alpha}^{2}}}-{\frac {2\,t\left({\frac {12\,\alpha\,x}{5}}+{t}^{2}\right)}{27\,{\alpha}^{3}}},
\end{aligned}
\end{eqnarray}
where the coefficients $P_{n}, Q_{n}$ and $R_{n}$ are obtained in following Theorem \ref{KBQthmKB2}.
\end{itemize}
\end{Case}

For similarity solutions of reductions corresponding to vector $V_{4}$ and $\alpha\,V_{1}+V_{3}$ we seek power series solution of the form
\begin{align}{\label{KBQ19}}
F = \sum_{n = 0}^{\infty}P_{n}\,\xi^{n},\;G = \sum_{n = 0}^{\infty}Q_{n}\,\xi^{n},\;H = \sum_{n = 0}^{\infty}R_{n}\,\xi^{n},
\end{align}
here $P_{n}, Q_{n}$ and $R_{n}$ are unknown coefficients of power series that need to be determined later.
On substituting \eqref{KBQ19} into reductions corresponding to respective vector fields we have following theorems:
\begin{thm}{\label{KBQthmKB1}}
\normalfont
Substitution of power series \eqref{KBQ19} into reductions corresponding to vector field $V_{4}$ gives following recurrence relations
\begin{equation}
\begin{aligned}
P_{{n+1}}=&-\,{\frac {10\,nQ_{{n}}+15\,\sum _{k=0}^{n} \left( n-k
 \right) P_{{k}}P_{{n-k}}+6\,\sum _{k=0}^{n}P_{{k}}P_{{n-k}}+8\,Q_{{n}
}}{6(n+1)}}\\
Q_{{n+1}}=&-\,{\frac {10\,nR_{{n}}+5\,\sum _{k=0}^{n} \left( n-k
 \right) P_{{k}}Q_{{n-k}}+10\,\sum _{k=0}^{n} \left( n-k \right) Q_{
{k}}P_{{n-k}}+8\,\sum _{k=0}^{n}P_{{k}}Q_{{n-k}}+12\,R_{{n}}}{6(n+1)}}\\
R_{{n+1}}=&\frac {-1}{108\,(n+1)}\Big[-125\,{n}^{3}P_{{n}}-375\,{n}^{2}P_{{n}}-330\,nP_{{
n}}+180\,\sum _{k=0}^{n}P_{{k}}R_{{n-k}}\\&+90\,\sum _{k=0}^{n} \left(
n-k \right) P_{{k}}R_{{n-k}}+180\,\sum _{k=0}^{n} \left( n-k
 \right) R_{{k}}P_{{n-k}}-80\,P_{{n}}\Big],\;\text{here}\; P_{0}, Q_{0}, R_{0}\;\text{ought to be taken as arbitrary.}
\end{aligned}
\end{equation}
and
\begin{align}
 P_{{1}}=-{P_{{0}}}^{2}-\frac{4\,Q_{{0}}}{3},Q_{{1}}=-\frac{4\,P_{{0}}Q_{{0
}}}{3}-2\,R_{{0}},R_{{1}}=-\frac{5\,P_{{0}}R_{{0}}}{3}+{\frac {20\,P_{{0}}}{27}}
\end{align}
\begin{proof}
For brevity we have omitted detailed calculations and results in form of power series solutions for system \eqref{KBQ1} corresponding to reductions under vector field $V_{4}$ are interpreted in Case \ref{KBQCaseKB2}.
\end{proof}
\end{thm}

\begin{thm}{\label{KBQthmKB2}}
\normalfont
Substitution of power series \eqref{KBQ19} into reductions corresponding to vector field $\alpha\,V_{1}+V_{3}$ gives following recurrence relations
\begin{equation}
\begin{aligned}
P_{{n+3}}=& \,\frac{-1}{1296\,{\alpha}^{4} \left( {n}^{3}+6\,{n}^{2}+11\,n+6 \right) }\Big[-450\,{\alpha}^{2}\sum _{k=0}^{n} \left( n-k+1
 \right) P_{{k}}R_{{n-k+1}}-900\,{\alpha}^{2}\sum _{k=0}^{n} \left(
n-k+1 \right) R_{{k}}P_{{n-k+1}}\\&+50\,\sum _{k=0}^{n}P_{{k}}P_{{n-k}}+
150\,\sum _{k=0}^{n} \left( n-k \right) P_{{k}}P_{{n-k}}-125\,Q_{{n}
}\Big],\\
Q_{{n+1}}=&\,\frac{-3}{2\,({n+1})}\, {\sum _{k=0}^{n} \left( n-k+1 \right) P_{{k}}
P_{{n-k+1}}},\\
R_{{n+1}}=&\,\frac{-1}{18\,{{\alpha}^{2}
 \left( n+1 \right) }}\, \Big[ 9\,{\alpha}^{2}\sum _{k=0}^{n} \left( n-k+1
 \right) P_{{k}}Q_{{n-k+1}}+18\,{\alpha}^{2}\sum _{k=0}^{n} \left( n
-k+1 \right) Q_{{k}}P_{{n-k+1}}+4\,nP_{{n}}+6\,P_{{n}}\Big]
\end{aligned}
\end{equation}
and
\begin{equation}
\begin{aligned}
 P_{{3}}=&\,{\frac {1350\,{\alpha}^{2}{P_{{0}}}^{3}P_{{1}}-1800\,
{\alpha}^{2}P_{{0}}P_{{1}}Q_{{0}}+3600\,{\alpha}^{2}P_{{1}}R_{{0}}-875
\,{P_{{0}}}^{2}+500\,Q_{{0}}}{31104\,{\alpha}^{4}}},\\
Q_{{1}}=&\,{\frac {1-18\,{\alpha}^{2}P_{{0}}P_{{1}}}{{12\alpha}^{2}}},\\
R_{{1}}=&\,{
\frac {6\,{\alpha}^{2}{P_{{0}}}^{2}P_{{1}}-8\,{\alpha}^{2}P_{{1}}Q_{{0
}}-3\,P_{{0}}}{{8\alpha}^{2}}},\\
 P_{{4}}=&\,\frac{1}{{186624\,{\alpha}^{6}}}\big[ 4050\,{\alpha}^{4}{P_{{0}}}^{3}P_{{2}}+2025\,
{\alpha}^{4}{P_{{0}}}^{2}{P_{{1}}}^{2}-4050\,{\alpha}^{4}P_{{0}}P_{{1}
}Q_{{1}}-5400\,{\alpha}^{4}P_{{0}}P_{{2}}Q_{{0}}+8100\,{\alpha}^{4}P_{
{1}}R_{{1}}\\&+10800\,{\alpha}^{4}P_{{2}}R_{{0}}-3000\,{\alpha}^{2}P_{{0}
}P_{{1}}+750\,{\alpha}^{2}Q_{{1}}+250\big],\\
Q_{{2}}=&\,
-\frac{3\,P_{{0}}P_{{2}}}{2}-\frac{3\,{P_{{1}}}^{2}}{4},\\
R_{{2}}=&\,{\frac {54\,{\alpha}^{
2}{P_{{0}}}^{2}P_{{2}}+27\,{\alpha}^{2}P_{{0}}{P_{{1}}}^{2}-54\,{
\alpha}^{2}P_{{1}}Q_{{1}}-72\,{\alpha}^{2}P_{{2}}Q_{{0}}-20\,P_{{1}}}{
72\,{\alpha}^{2}}},
\end{aligned}
\end{equation}
here $P_{0}, Q_{0}, R_{0}, P_{1}, P_{2}$ ought to be taken as arbitrary.
\begin{proof}
Again for brevity we have omitted detailed calculations and results in form of power series solutions for system \eqref{KBQ1} corresponding to reductions under vector field $\alpha\,V_{1}+V_{3}$ are interpreted in Case \ref{KBQCaseKB3}.
\end{proof}
\end{thm}
\section{Conservation laws}{\label{KBQsec3}}
In physics, the conservation laws are fundamental laws those ensures that the certain physical quantity will not change
with time during the course of physical process \cite{singh2016backlund}. Some of the well-known conservation laws in physics are conservation of
mass, momentum, energy and electric charge etc. It is well-known fact that the Noether's theorem gives conservation laws for a system only when it has variational principle. To establish conservation laws for a system without variational structure, Ibragimov  \cite{ibragimov2007new} has given a new theorem based on the concept of adjoint equations for nonlinear equations. In the recent literature, many authors have applied the theorem of Ibragimov to derive conservations laws. For instance, in Ref. \cite{ibragimov2011self} it was proved that the Camassa-Holm is strictly self-adjoint and conservation laws were also obtained without classical Lagrangians. In Ref. \cite{freire2011nonlinear} authors have constructed some
conservation laws for nonlinear self-adjoint class of the generalized fifth-order equation, such as a general Kawahara equation, modified
Kawahara equation and simplified modified Kawahara equation. Johnpillai and Khalique \cite{johnpillai2011conservation} has applied the same theorem to derive conservation laws for generalized KdV equation of time-dependent variable
coefficients. For further details about application of theorem by Ibragimov, see recent work \cite{torrisi2013quasi,wang2013symmetry,bozhkov2013conservation,bozhkov2013group,wang2015nonlocal,gandarias2016symmetries,abdulwahhab2016exact,gupta2017group}.\\
\indent Based on theory developed in \cite{ibragimov2007new} and notations adopted therein, we define formal Lagrangian for equations \eqref{KBQ1} in the following manner

\begin{align}
I = \label{KBQ20}\phi(x,t)\left(u_{t}-\frac{3}{2}uu_{x}-v_{x} \right)
+\psi(x,t)\left(v_{t}-vu_{x}-\frac{1}{2}uv_{x}-w_{x} \right)
+\theta(x,t)\left(w_{t}+\frac{1}{4}u_{xxx}-wu_{x}-\frac{1}{2}uw_{x}\right),
\end{align}
here $\phi(x,t), \psi(x,t),$ and $\theta(x,t)$ are new dependent variables. The adjoint equations for \eqref{KBQ1} can be written as
\begin{align}{\label{KBQ22}}
F^{*} = \frac{\delta I}{\delta u}=0, \;G^{*} = \frac{\delta I}{\delta v}=0,\;
H^{*} = \frac{\delta I}{\delta w}=0,
\end{align}
here we have used variational derivative $\frac{\delta }{\delta u^{\alpha}}$ defined by the relation
\begin{align}{\label{KBQ21}}
\frac{\delta }{\delta u^{\alpha}} = \frac{\partial}{\partial u^{\alpha}}+\sum_{s=1}^{\infty}(-1)^{s}D_{i_{1}}\dots D_{i_{s}}\frac{\partial}{\partial u^{\alpha}_{i_{1}\dots i_{s}}}.
\end{align}
Substituting Lagrangian \eqref{KBQ20} into \eqref{KBQ22} and using relation \eqref{KBQ21}, we obtain adjoint equations
\begin{equation}{\label{KBQ26}}
\begin{aligned}
F^{*} = &\frac{1}{2}\,\psi v_{x}+\frac{1}{2}\,\theta w_{x}+\frac{3}{2}\,\phi_{x}u+\psi_{x}v+
\theta_{x}w-\theta_{t}-\frac{1}{4}\,\theta_{xxx}=0,\\
G^{*} = & -\frac{1}{2}\,\psi u_{x}+\phi_{x}+\frac{1}{2}\,\psi_{x}u-\psi_{t}=0,\\
H^{*} = &-\frac{1}{2}\,\theta u_{x}+\psi_{x}+\frac{1}{2}\,\theta_{x}u-\theta_{t}=0.
\end{aligned}
\end{equation}
 For conservation laws we shall use following theorem proved in \cite{ibragimov2007new}.
 \begin{thm}
 \normalfont
 Any infinitesimal symmetry (\emph{Lie point, Lie B$\ddot{ a}cklund$, nonlocal})
 \begin{align*}
 V = \xi^{i}(x, u, u_{(1)},\dots)\,\frac{\partial}{\partial x^{i}}+\eta^{\alpha}(x, u, u_{(1)},\dots)\frac{\partial}{\partial u^{\alpha}}
 \end{align*}
 of equations \eqref{KBQ1} leads to conservation laws $D_{i}(C^{i}) = 0$ constructed by formula
 \begin{equation}{\label{KBQ23}}
 \begin{aligned}
 C^{i} = \xi^{i}I+&W^{\alpha}\left[\frac{\partial I}{\partial u_{i}^{\alpha}}-D_{j}\left(\frac{\partial I}{\partial u_{ij}^{\alpha}}\right)+D_{j}D_{k}\left(\frac{\partial I}{\partial u_{ijk}^{\alpha}}\right)-\dots\right]\\
 +&D_{j}(W^{\alpha})\left[\frac{\partial I}{\partial u_{ij}^{\alpha}}-D_{k}\left(\frac{\partial I}{\partial u_{ijk}^{\alpha}}\right)+\dots\right]+D_{j}D_{k}(W^{\alpha})\left[\frac{\partial I}{\partial u_{ijk}^{\alpha}}-\dots\right],
 \end{aligned}
 \end{equation}
 here $W^{\alpha} = \eta^{\alpha}-\xi^{j}u_{j}^{\alpha}$ and $I$ is Lagrangian defined by \eqref{KBQ20}.
 \end{thm}
 The relation \eqref{KBQ23} can be simplified by writing Lagrangian $I$ with respect to all mixed derivative $u_{ij}^{\alpha}, u_{ijk}^{\alpha},\dots$ in symmetric manner. We obtain
 \begin{equation}{\label{KBQ24}}
 \begin{aligned}
 C^{x} = \,&\xi\,I+W^{(1)}\left[\frac{\partial I}{\partial u_{x}}-D_{x}\left(\frac{\partial I}{\partial u_{xx}}\right)+D_{x}^{2}\left(\frac{\partial I}{\partial u_{xxx}}\right)\right]+
 W^{(2)}\,\frac{\partial I}{\partial v_{x}}+W^{(3)}\,\frac{\partial I}{\partial w_{x}}\\
 &+D_{x}(W^{(1)})\left[\left(\frac{\partial I}{\partial u_{xx}}\right)-D_{x}\left(\frac{\partial I}{\partial u_{xxx}}\right)\right]+D_{x}^{2}(W^{(1)})\,\frac{\partial I}{\partial u_{xxx}},\\
 C^{t} = \,&\tau \,I+W^{(1)}\frac{\partial I}{\partial u_{t}}+W^{(2)}\frac{\partial I}{\partial v_{t}}+W^{(3)}\frac{\partial I}{\partial w_{t}},
 \end{aligned}
 \end{equation}
 here $D_{i}$ denotes operator of total differentiation:
 \begin{align*}
 D_{i} = \frac{\partial}{\partial x^{i}}+u_{i}^{\alpha}\frac{\partial}{\partial u^{\alpha}}+
 u_{ij}^{\alpha}\frac{\partial}{\partial u_{j}^{\alpha}}+\dots,
 \end{align*}
 and rest of details about notations can be seen in Ref. \cite{ibragimov2007new}.\\
 In following cases we shall find conserved currents \eqref{KBQ24} corresponding to every symmetry generator of optimal system obtained in Theorem \ref{KBQthm}.
 \begin{case}
 \normalfont
 For generator $V_{1} = \frac{\partial}{\partial t}$, the Lie's characteristic functions are obtained as follow
 \begin{align}
 \label{KBQ27}W^{(1)} = -u_{t},\;W^{(2)} = -v_{t},\;W^{(3)} = -w_{t}.
 \end{align}
 Substituting \eqref{KBQ27} into \eqref{KBQ24} yield the following conserved currents

 \begin{equation}{\label{KBQ50}}
 \begin{aligned}
 C^{x} = &\,\frac{3}{2}\,u_{{t}}\phi\,u+u_{{t}}\psi\,v+u_{{t}}\theta\,w-\frac{1}{4}\,u_{{t}}\theta
_{{xx}}+v_{{t}}\phi+\frac{1}{2}\,v_{{t}}\psi\,u+w_{{t}}\psi+\frac{1}{2}\,w_{{t}}
\theta\,u+\frac{1}{4}\,u_{{xt}}\theta_{{x}}-\frac{1}{4}\,u_{{xxt}}\theta,\\
 C^{t} = &\,-\frac{3}{2}\,\phi\,uu_{{x}}-\phi\,v_{{x}}-\psi\,vu_{{x}}-\frac{1}{2}\,\psi\,uv_{{x}}-
\psi\,w_{{x}}+\frac{1}{4}\,\theta\,u_{{xxx}}-\theta\,wu_{{x}}-\frac{1}{2}\,\theta\,u
w_{{x}},
 \end{aligned}
 \end{equation}
 where $\phi(x,t), \psi(x,t)$ and $\theta(x,t)$ are arbitrary solutions of adjoint equations \eqref{KBQ26}.
 \end{case}

 \begin{case}
 \normalfont
 For generator $V_{2} = \frac{\partial}{\partial x}$, the Lie's characteristic functions are obtained as follow
 \begin{align}
 \label{KBQ25}W^{(1)} = -u_{x},\;W^{(2)} = -v_{x},\;W^{(3)} = -w_{x}.
 \end{align}
 Substituting \eqref{KBQ25} into \eqref{KBQ24} yield the following conserved currents
 \begin{equation}{\label{KBQ49}}
 \begin{aligned}
 C^{x} = &\,u_{{t}}\phi+v_{{t}}\psi+w_{{t}}\theta-\frac{1}{4}\,u_{{x}}\theta_{{xx}}+\frac{1}{4}\,
u_{{x,x}}\theta_{{x}},\\
C^{t} = &\,-\phi\,u_{{x}}-\psi\,v_{{x}}-\theta\,w_{{x}},
 \end{aligned}
 \end{equation}
 where $\phi(x,t), \psi(x,t)$ and $\theta(x,t)$ are arbitrary solutions of adjoint equations \eqref{KBQ26}.
 \end{case}
 \begin{case}
 \normalfont
 For generator $V_{3} = -\frac{5t}{6}\frac{\partial}{\partial x}+\frac{\partial}{\partial u}-\frac{2u}{3}\frac{\partial}{\partial v}-\frac{v}{3}\frac{\partial}{\partial w}$, the Lie's characteristic functions are obtained as follow
 \begin{align}
 \label{KBQ28}W^{(1)} = 1+\frac{5\,t}{6}u_{x},\;W^{(2)} = -\frac{2\,u}{3}+\frac{5\,t}{6}v_{x},\;W^{(3)} = -\frac{v}{3}+\frac{5\,t}{6}w_{x}.
 \end{align}
 Substituting \eqref{KBQ28} into \eqref{KBQ24} yield the following conserved currents

 \begin{equation}{\label{KBQ48}}
 \begin{aligned}
 C^{x} = &\,\frac{1}{3}\,\psi\,{u}^{2}-\frac{5}{6}\,tu_{{t}}\phi-\frac{5}{6}\,tv_{{t}}\psi-\frac{5}{6}\,tw_{{t}}
\theta+{\frac {5\,tu_{{x}}\theta_{{xx}}}{24}}+\frac{1}{6}\,v\theta\,u-{\frac
{5\,tu_{{xx}}\theta_{{x}}}{24}}\\
&+\frac{1}{4}\,\theta_{{xx}}-\frac{2}{3}\,\psi\,v-
\theta\,w-\frac{5}{6}\,\phi\,u
,\\
C^{t} = &\,\phi+\frac{5}{6}\,\phi\,tu_{{x}}-\frac{2}{3}\,\psi\,u+\frac{5}{6}\,\psi\,tv_{{x}}-\frac{1}{3}\,v\theta
+\frac{5}{6}\,\theta\,tw_{{x}}
,
 \end{aligned}
 \end{equation}
 where $\phi(x,t), \psi(x,t)$ and $\theta(x,t)$ are arbitrary solutions of adjoint equations \eqref{KBQ26}.
 \end{case}

 \begin{case}
 \normalfont
 For generator $V_{4} = \frac{3x}{5}\frac{\partial}{\partial x}+t\frac{\partial}{\partial t}-\frac{2u}{5}\frac{\partial}{\partial u}-\frac{4v}{5}\frac{\partial}{\partial v}-\frac{6w}{5}\frac{\partial}{\partial w}$, the Lie's characteristic functions are obtained as follow
 \begin{align}
 \label{KBQ29}W^{(1)} = -\frac{2\,u}{5}-\frac{3\,x}{5}u_{x}-tu_{t},\;W^{(2)} = -\frac{4\,v}{5}-\frac{3\,x}{5}v_{x}-tv_{t},\;W^{(3)} = -\frac{6\,w}{5}-\frac{3\,x}{5}w_{x}-tw_{t}.
 \end{align}
 Substituting \eqref{KBQ29} into \eqref{KBQ24} yield the following conserved currents

 \begin{equation}{\label{KBQ47}}
 \begin{aligned}
 C^{x} = &\,\frac{3}{5}\,\phi\,{u}^{2}-\frac{1}{10}\,u\theta_{{xx}}+\frac{4}{5}\,v\phi+\frac{6}{5}\,w\psi+\frac{1}{4}\,
\theta_{{x}}u_{{x}}-\frac{2}{5}\,\theta\,u_{{xx}}+\frac{3}{2}\,tu_{{t}}\phi\,u+tu_{{t
}}\psi\,v+tu_{{t}}\theta\,w+\frac{1}{2}\,tv_{{t}}\psi\,u+\frac{1}{2}\,tw_{{t}}\theta\,
u\\
&+tv_{{t}}\phi+u\theta\,w-\frac{1}{4}\,tu_{{t}}\theta_{{xx}}+{\frac {3\,
\theta_{{x}}xu_{{xx}}}{20}}+\frac{1}{4}\,\theta_{{x}}tu_{{xt}}-\frac{1}{4}\,\theta\,
tu_{{xxt}}+\frac{4}{5}\,u\psi\,v\\
&+\frac{3}{5}\,x\psi\,v_{{t}}+\frac{3}{5}\,x\phi\,u_{{t}}-{
\frac {3\,xu_{{x}}\theta_{{xx}}}{20}}+\frac{3}{5}\,x\theta\,w_{{t}}+tw_{{t}}
\psi
\\
C^{t} = &\,-\frac{3}{2}\,t\phi\,uu_{{x}}-t\phi\,v_{{x}}-t\psi\,vu_{{x}}-\frac{1}{2}\,t\psi\,uv_{{
x}}-t\psi\,w_{{x}}+\frac{1}{4}\,tu_{{xxx}}\theta-t\theta\,wu_{{x}}-\frac{1}{2}\,t
\theta\,uw_{{x}}-\frac{2}{5}\,\phi\,u-\frac{3}{5}\,\phi\,xu_{{x}}\\
&-\frac{4}{5}\,\psi\,v-\frac{3}{5}\,
\psi\,xv_{{x}}-\frac{6}{5}\,\theta\,w-\frac{3}{5}\,\theta\,xw_{{x}},
 \end{aligned}
 \end{equation}
 where $\phi(x,t), \psi(x,t)$ and $\theta(x,t)$ are arbitrary solutions of adjoint equations \eqref{KBQ26}.
 \end{case}

 \begin{case}
 \normalfont
 For generator $\alpha\,V_{1}+V_{3} = -\frac{5t}{6}\frac{\partial}{\partial x}+\alpha\,\frac{\partial}{\partial t}+\frac{\partial}{\partial u}-\frac{2u}{3}\frac{\partial}{\partial v}-\frac{v}{3}\frac{\partial}{\partial w}$, the Lie's characteristic functions are obtained as follow
 \begin{align}
 \label{KBQ30}W^{(1)} = 1+\frac{5\,t}{6}u_{x}-\alpha\,u_{t},\;W^{(2)} = -\frac{2\,u}{3}+\frac{5\,t}{6}v_{x}-\alpha\,v_{t},\;W^{(3)} = -\frac{v}{3}+\frac{5\,t}{6}w_{x}-\alpha\,w_{t}.
 \end{align}
 Substituting \eqref{KBQ30} into \eqref{KBQ24} yield the following conserved currents

 \begin{equation}\label{KBQ46}
 \begin{aligned}
 C^{x}=&\,\frac{3}{2}\,\alpha\,u_{{t}}\phi\,u+\alpha\,u_{{t}}\psi\,v+\alpha\,u_{{t}}
\theta\,w+\frac{1}{2}\,\alpha\,v_{{t}}\psi\,u+\frac{1}{2}\,\alpha\,w_{{t}}\theta\,u+\frac{1}{6}\,v\theta\,u+\frac{1}{4}\,\theta_{{x}}\alpha\,u_{{xt}}+{\frac {5\,tu_{{x}}
\theta_{{xx}}}{24}}-\frac{5}{6}\,tw_{{t}}\theta\\
&+\alpha\,v_{{t}}\phi-\frac{1}{4}\,
\alpha\,u_{{t}}\theta_{{xx}}-\frac{1}{4}\,\theta\,\alpha\,u_{{xxt}}-\frac{5}{6}\,tv
_{{t}}\psi-\frac{5}{6}\,tu_{{t}}\phi-{\frac {5\,tu_{{xx}}\theta_{{x}}}{24}}+
\alpha\,w_{{t}}\psi\\
&+\frac{1}{4}\,\theta_{{xx}}-\frac{5}{6}\,\phi\,u-\frac{2}{3}\,\psi\,v
-
\theta\,w+\frac{1}{3}\,\psi\,{u}^{2},\\
C^{t} = &\,-\frac{3}{2}\,\alpha\,\phi\,uu_{{x}}-\alpha\,\phi\,v_{{x}}-\alpha\,\psi\,vu_{{
x}}-\frac{1}{2}\,\alpha\,\psi\,uv_{{x}}-\alpha\,\psi\,w_{{x}}+\frac{1}{4}\,\alpha\,
\theta\,u_{{xxx}}-\alpha\,\theta\,wu_{{x}}-\frac{1}{2}\,\alpha\,\theta\,uw_{
{x}}+\phi+\frac{5}{6}\,\phi\,tu_{{x}}\\
&-\frac{2}{3}\,\psi\,u+\frac{5}{6}\,\psi\,tv_{{x}}-\frac{1}{3}\,v
\theta+\frac{5}{6}\,\theta\,tw_{{x}} ,
 \end{aligned}
 \end{equation}
 where $\phi(x,t), \psi(x,t)$ and $\theta(x,t)$ are arbitrary solutions of adjoint equations \eqref{KBQ26}. In similar manner, the conserved currents corresponding to generator $\alpha\,V_{1}-V_{3} = \frac{5t}{6}\frac{\partial}{\partial x}+\alpha\,\frac{\partial}{\partial t}-\frac{\partial}{\partial u}+\frac{2u}{3}\frac{\partial}{\partial v}+\frac{v}{3}\frac{\partial}{\partial w}$ can also be calculated.
 \end{case}
 \begin{rem}
 \normalfont
 Despite the huge success of new conservation theorem of Ibragimov, the recent comments from Stephen C. Anco \cite{anco2016incompleteness} confirms the incompleteness of the theorem. In particular, the formulation proposed by Ibragimov can generate trivial conservation laws and does
not always yield all non-trivial conservation laws. But fortunately in present case, all the conservation laws given at  \eqref{KBQ50}, \eqref{KBQ49}, \eqref{KBQ48},\eqref{KBQ47} and \eqref{KBQ46} are all non-trivial one.
 \end{rem}
\section{Conclusion}{\label{KBQsec4}}
Using classical Lie symmetry analysis we have analyzed three fields Kaup-Boussinesq system \eqref{KBQ1} in a comprehensive manner. Based on Killing's form derived in Lemma \ref{KBQlem}, the complete classification of Lie algebra \eqref{KBQ5} is obtained in Theorem \ref{KBQthm}.  Similarity reductions and invariant solutions using power series method are also presented. Apart from this usual symmetry analysis,  we have demonstrated the construction of several nonlocal conservation laws based on the theory of a new conservation
theorem \cite{ibragimov2007new}. The work presented here emphasized the relevance of new conservation theorem by Ibragimov for construction of conservation from Lie symmetries without the formulation of classical Lagrangian. 
\section*{Acknowledgements}
Rajesh Kumar Gupta thanks the University Grant Commission for sponsoring this research under Research Award Scheme (F. 30-105/2016 (SA-II)).
%\bibliography{My.Bibtex.Library}
%   \bibliographystyle{elsarticle-num}
   
\end{document}